\documentclass{article}
\usepackage{ijcai24}

\usepackage{times}
\usepackage{soul}
\usepackage{url}
\usepackage[hidelinks]{hyperref}
\usepackage[utf8]{inputenc}
\usepackage[small]{caption}
\usepackage{graphicx}
\usepackage{amsmath}
\usepackage{amsthm}
\usepackage{booktabs}
\usepackage{algorithm}
\usepackage{algorithmic}
\usepackage[switch]{lineno}
\usepackage{subfigure}

\DeclareMathOperator*{\argmax}{arg\,max} % Jan Hlavacek
\title{Hedging Beyond the Mean: A Distributional Reinforcement Learning Perspective for Hedging Portfolios with Structured Products}

\author{
Anil Sharma$^{1}$
\and
Freeman Chen$^{2}$\and
Jaesun Noh$^2$\and %$^{2,3}$\and
Julio DeJesus$^2$\And
Mario Schlener$^2$
\affiliations
$^1$Ernst \& Young LLP, India, %\\
$^2$Ernst \& Young - Canada %\\
\emails
anil.sharma2@in.ey.com,
\{freeman.chen, jaesun.noh, julio.dejesus, mario.schlener\}@ca.ey.com
}

\begin{document}

\maketitle

\begin{abstract}
    Research in quantitative finance has demonstrated that reinforcement learning (RL) methods have delivered promising outcomes in the context of hedging financial portfolios. For example, hedging a portfolio of European options using RL achieves better $PnL$ distribution than the trading hedging strategies like Delta neutral and Delta-Gamma neutral~\cite{2q_Cao_2020}. There is great attention given to the hedging of vanilla options, however, very little is mentioned on hedging a portfolio of structured products such as Autocallable notes. Hedging structured products is much more complex and the traditional RL approaches tend to fail in this context due to the underlying complexity of these products. These are more complicated due to presence of several barriers and coupon payments, and having a longer maturity date (from $7$ years to a decade), etc. In this direction, we propose a distributional RL based method to hedge a portfolio containing an Autocallable structured note. We will demonstrate our RL hedging strategy using American and Digital options as hedging instruments. Through several empirical analysis, we will show that distributional RL provides better $PnL$ distribution than traditional approaches and learns a better policy depicting lower value-at-risk ($VaR$) and conditional value-at-risk ($CVaR$), showcasing the potential for enhanced risk management.
\end{abstract}

\section{Introduction}
\label{sec:intro}

In the dynamic landscape of financial markets, portfolio managers are faced with the challenge of managing risks associated with their investment holdings. These portfolio managers employ several hedging strategies to minimize the associated risks, for example, using Delta-Gamma neutral hedging strategies to create a Gamma neutral portfolio. The portfolio manager needs to not only identify suitable hedging instruments but also determine the optimal hedging ratios to effectively mitigate these risks. Moreover, the dynamic nature of financial markets demands a hedging strategy that can adapt in real-time to evolving market conditions.
% A client portfolio is dynamic and evolving, as the portfolio manager may create short or long positions regularly. 
Traditional approaches to hedging do not provide the flexibility and adaptability required for such dynamic decision-making. However, the emergence of Reinforcement Learning (RL) presents a promising avenue for addressing these challenges.% RL, a subfield of artificial intelligence, empowers decision-makers to make sequential choices in an environment without relying on predefined rules or prior knowledge. 

Researchers have demonstrated that RL is an attractive alternative to traditional hedging strategies based on the performance and Profit and Loss ($PnL$) distribution as compared to the baseline Delta neutral and Delta-Gamma neutral strategies~\cite{2q_Cao_2020,kolm_ritter,cao2023_gamma_vega}. When undertaking the hedging of a portfolio containing vanilla options (eg., European and American),  better $PnL$ was achieved than these traditional hedging strategies. Researchers have greatly focused on using RL for hedging of European options~\cite{cao2023_gamma_vega,2q_Cao_2020,kolm_ritter} and ~\cite{chen2023hedging_barrier} have proposed a new RL based hedging method for Barrier options. %However, there is a little on hedging of other very risky options such as Autocallable notes, which are complicated due to their structure and longer maturity.
Recently, \cite{autocall_hedge_mc} proposed a method for pricing Autocallable notes and employing a Delta neutral strategy for hedging. They emphasized that hedging Autocallable notes is complex due to their intricate structure. Similarly, we investigate Autocallable structured note hedging using RL. %As highlighted in \cite{autocall_hedge_mc}, the hedging of Autocallable notes is intricate due to its complex structure. They further demonstrate that the Delta of the portfolio becomes very large near the barriers (call dates), resulting in a substantial increase in the cost of the hedging task. In the same direction, we explore hedging of the Autocallable structured note using RL. 
To the best of our knowledge, we are the first in the public domain to investigate the hedging of structured products using RL. We propose a new RL based method to hedge a portfolio containing a short Autocallable note, where we achieve better $PnL$ distribution than traditional methods. %We use Autocallable note structure of Canadian Imperial Bank of Commerce. 

\begin{figure}
\centering
    \includegraphics[width=7.8cm]{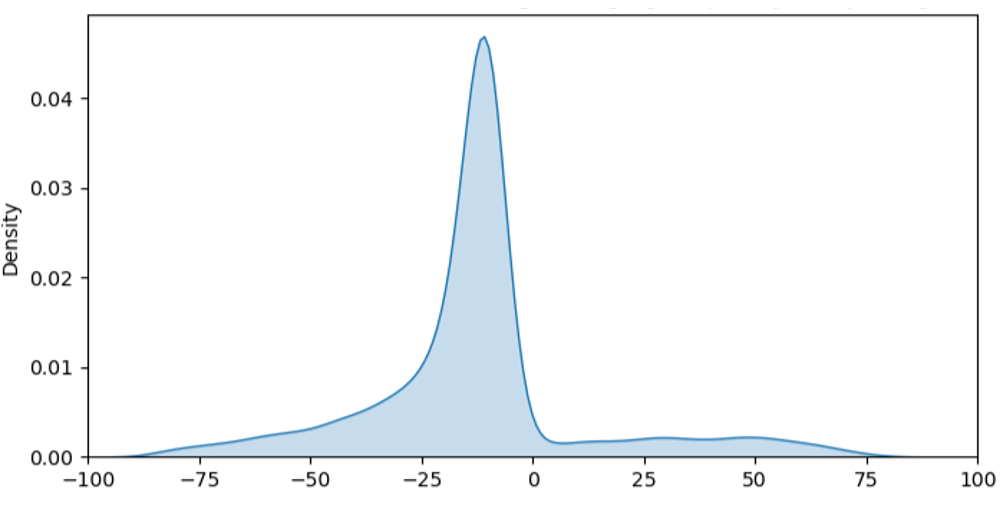}
\caption{The $PnL$ distribution when no hedging is performed.}
\label{fig:pnl_no_hedge}
\end{figure}

% In this context, we explore the application of RL algorithms to derive an optimal hedging strategy for a portfolio manager overseeing Autocallable notes. 
Unlike vanilla options, Autocallable notes are a structured product having an embedded exotic nature through several barriers and coupon payments. These are more complicated than vanilla options and have longer maturity (for example, $7$ years to a decade). Due to this complicated structure, hedging of Autocallable notes is a complex task. Hence, Autocallable notes, with their inherent exposure to market fluctuations, require careful risk management to safeguard against adverse movements in the portfolio Gamma. In this paper, we investigate an Autocallable note structure with the underlying as a simulated Geometric Brownian Motion (GBM) process with a SABR volatility model. We propose a hedging method using RL to hedge a trader's portfolio having a short Autocallable note. The probability distribution of the unhedged $PnL$ for vanilla options is highly symmetrical and hence helps to learn an optimal policy when used as a reward function in the RL algorithm. Also, ~\cite{cao2023_gamma_vega} demonstrated that the distributional RL method achieves better $PnL$ for European options as compared to the classical RL methods (classical RL uses expected reward as a learning criteria). We extend their approach to American options with early exercise using the Longstaff Schwartz algorithm~\cite{schwartz2001valuing_lsmc} which helps to address the complexities introduced by the option's early exercise feature. In contrast, for the Autocallable note with a maturity of 7 years, the $PnL$ distribution is characterized by significant skewness, as illustrated in Figure \ref{fig:pnl_no_hedge}. This skewness is due to the several coupon payments after a monthly call barrier. Therefore, hedging such a portfolio is more complicated due to the skewed reward function. In this paper, we demonstrate learning a policy for hedging under this skewed reward distribution. 

% The unhedged $PnL$ distribution of the vanilla options is highly symmetric whereas for the Autocallable note, for a maturity of $7$ years, it is highly skewed as shown in Figure~\ref{fig:pnl_no_hedge} and hence hedging it become more complicated with traditional approaches. We demonstrate our modifications to hedge this skewed distribution. 

% ~\cite{cao2023_gamma_vega} demonstrated that the distributional RL method produces better results on European options as compared to the traditional RL methods. They used Distributed Distributional Deep Deterministic Policy Gradients (D4PG) to learn action distribution. We extend their approach to American options with early exercise using the Longstaff Schwartz algorithm as proposed in~\cite{schwartz2001valuing_lsmc}. 

In our approach, we use Distributed Distributional DDPG algorithm with Quantile Regression (QR) to learn an optimal policy for hedging. The distributional RL enables a more nuanced understanding of uncertainty and risk in the learning process. Specifically, we use a Digital/American option as hedging instruments and the trained RL agent selects an action which quantifies the amount of hedging that needs to be performed. We compare the $PnL$ distribution, Value at Risk ($VaR$), and Conditional Value at Risk ($CVaR$) of different hedging strategies and show that the RL algorithm not only reduces $95\%VaR$, but also makes the $PnL$ distribution more symmetric and retains positive returns.

Our specific contributions are the following: 
\begin{enumerate}
    \item We propose a distributional RL based method to hedge a portfolio containing one short Autocallable note. We conduct a thorough analysis and introduce a novel objective function which helps to learn a generalized policy for several portfolio parameters. 
    \item We demonstrate the intricacies associated with hedging structured products, highlighting the inadequacies of classical RL approaches, particularly in the context of the skewed unhedged $PnL$ distribution. %We incorporate various modifications, including the inclusion of Delta exposure, to enhance the hedging strategy. 
    % We demonstrate the intricacies associated with hedging structured products, highlighting the inadequacies of conventional RL approaches, particularly in the context of the skewed unhedged $PnL$ distribution. Our approach incorporates various modifications, including the inclusion of Delta exposure, to enhance the hedging strategy. 
    \item We compare with traditional hedging strategies, including Delta neutral and Delta-Gamma neutral.
    
\end{enumerate}

\section{Related Works}
\label{sec:related}

Recently, Reinforcement Learning (RL) based methods have been explored extensively in finance, for example, optimal trade execution~\cite{zhang2023generalizable_optimaltrade}, credit pricing~\cite{khraishi2022offline_creditpricing_icaif22}, market making~\cite{ganesh2019reinforcement_marketmaking}, learning exercise policies for American options~\cite{pmlr-v5-li09d_earlyexerciseRL}, optimal hedging~\cite{2q_Cao_2020,kolm_ritter,deeper_hedging_Gao_2023,murray2022deep_icaif22}, etc.  

Hedging a portfolio of exotic options is a fundamental research problem which requires sequential decision making to re-balance the portfolio at different intervals for hedging. This sequential decision making has attracted reinforcement learning based solutions which outperform traditional hedging strategies. For example, \cite{kolm_ritter} have used Deep Q-learning and Proximal Policy Optimization (PPO) algorithms in RL to learn a policy for option replications with market frictions. They learn a single policy which works with several strike prices. \cite{2021_giurca_delta} used RL for delta hedging under various parameters such as transaction cost, option maturity, and hedging frequency. They applied transfer learning to use the policy trained on simulated data for real data. \cite{2q_Cao_2020} and \cite{deeper_hedging_Gao_2023} used DDPG (Deep Deterministic Policy Gradient) RL algorithm to perform hedging under SABR and Heston volatility models respectively. 

\cite{2022_empirical_deep_hedging} trained an RL algorithm directly on real data for intraday options, over 6 years, on the $S\&P500$ index. \cite{jpmorgan_market_friction} hedged over-the-counter derivatives using RL algorithm under market frictions like trading cost and liquidity constraints. \cite{cva_hedge} used reinforcement learning for CVA hedging. The literature has given much attention on using RL based sequential decision making on several European~\cite{kolm_ritter,2q_Cao_2020,halperin2019qlbs}, American~\cite{pmlr-v5-li09d_earlyexerciseRL}, and Barrier options~\cite{chen2023hedging_barrier}. These methods have demonstrated that RL is an attractive alternative to traditional hedging strategies based on the performance and $PnL$ distribution. However, there is little attention on hedging very risky options such as Autocallable notes. Recently, \cite{autocall_hedge_mc} proposed a method for pricing Autocallable notes and employing a Delta neutral strategy for hedging. They emphasized that the hedging is very complicated due to presence of several barriers, noting the high cost of Delta hedging near these barriers. In contrast, our approach involves training a RL policy for hedging of a 7-year maturity Autocallable note portfolio. % Recently, \cite{autocall_hedge_mc} has proposed a method for the pricing of autocallable notes and hedging using Delta neutral strategy. As our observation, they emphasized that the hedging is very complicated due to presence of several barriers. They showed that Delta hedging is becomes very costly because the Delta value of the portfolio near the barrier is very high. Whereas, we learn a policy using RL to hedge such a portfolio of Autocallable note with a maturity of $7$ years. 
We show the performance of an RL agent under varying transaction cost and choosing different options as hedging instruments. We demonstrate that the RL agent provides better $PnL$ distribution than traditional hedging strategies. 

\section{Problem Formulation}
\label{sec:problem}

In this section, we will describe the simulation which generates the asset and the Autocallable note price, followed by the problem formulation using Markov Decision Process (MDP). 

\subsection{The Asset and Option Pricing}
In actual examples, traders have extensively used the SABR volatility model~\cite{2q_Cao_2020,cao2023_gamma_vega} to simulate the stock paths with volatility and to perform automated trading/hedging decisions. The SABR model, named (Stochastic Alpha, Beta, Rho), is a widely employed framework %that incorporates parameters such as alpha, beta, and rho 
to capture different aspects of the asset volatility. In this paper, we also utilize the SABR volatility model in our simulation, assuming the underlying stock price $x$ follows a Geometric Brownian motion (GBM) with SABR volatility.

\begin{table}[t]
\caption{The details of the Autocallable coupon note structure on U.S. Select Regional Banks Index (AR).}
\label{tab:autocall_str}
\centering
 \begin{tabular}{|c | c|}
\hline
 Variable & Value \\ \hline
 Reference Index  & Solactive United States Select\\ & Regional Bank Index AR \\ \hline
 % Reference Index  & Any stock price \\ \hline
 Initial Price  & \$100 \\ \hline
 Term (0\% fee)  & $7$ years \\ \hline
 Coupon Frequency & Monthly  \\ \hline
 Coupon Rate  & 0.95\%  \\ \hline
 Coupon Barrier  & -35.00\%  \\ \hline
 Autocall Frequency  & Semi-Annual \\ \hline
 Call Barrier  & 0.00\% \\ \hline
 Contingent Principal & \\ Protection  & -35.00\%  \\ \hline
     
 \end{tabular}
\end{table}

\begin{figure}
\centering
    \includegraphics[width=8cm]{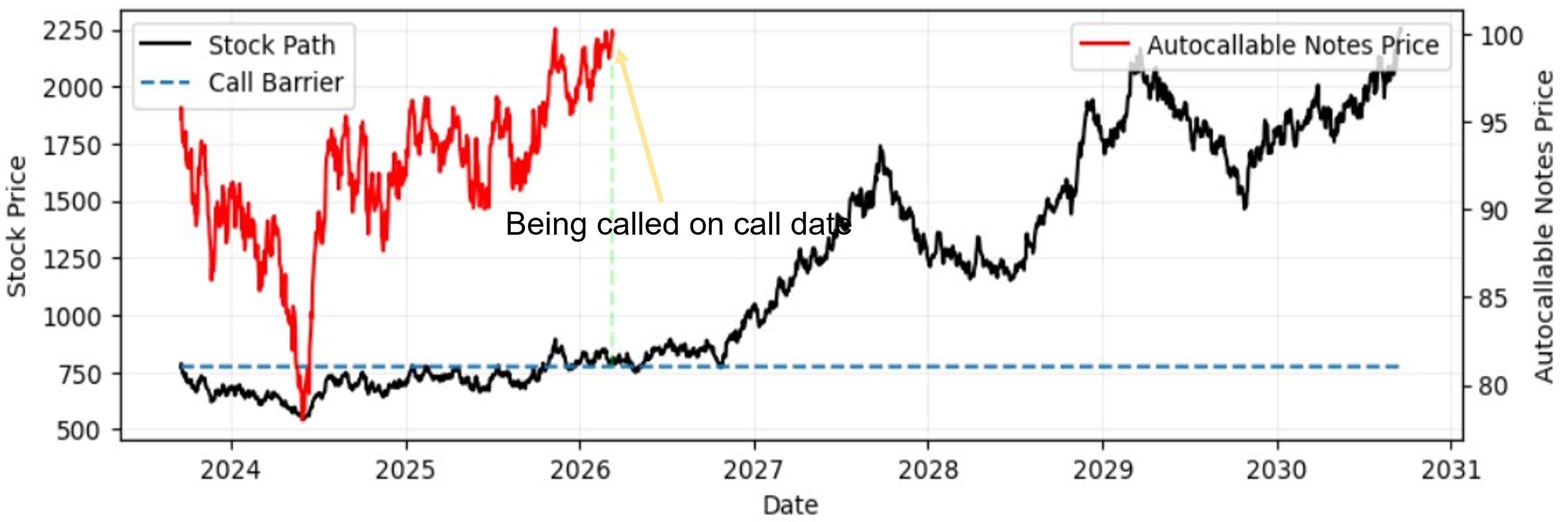}
\caption{The simulated stock path and the Autocallable note price being called early after $2.5$ years.}
\label{fig:early_call_autocallable}
\end{figure}

\begin{figure*}
\begin{center}
    \includegraphics[width=15cm]{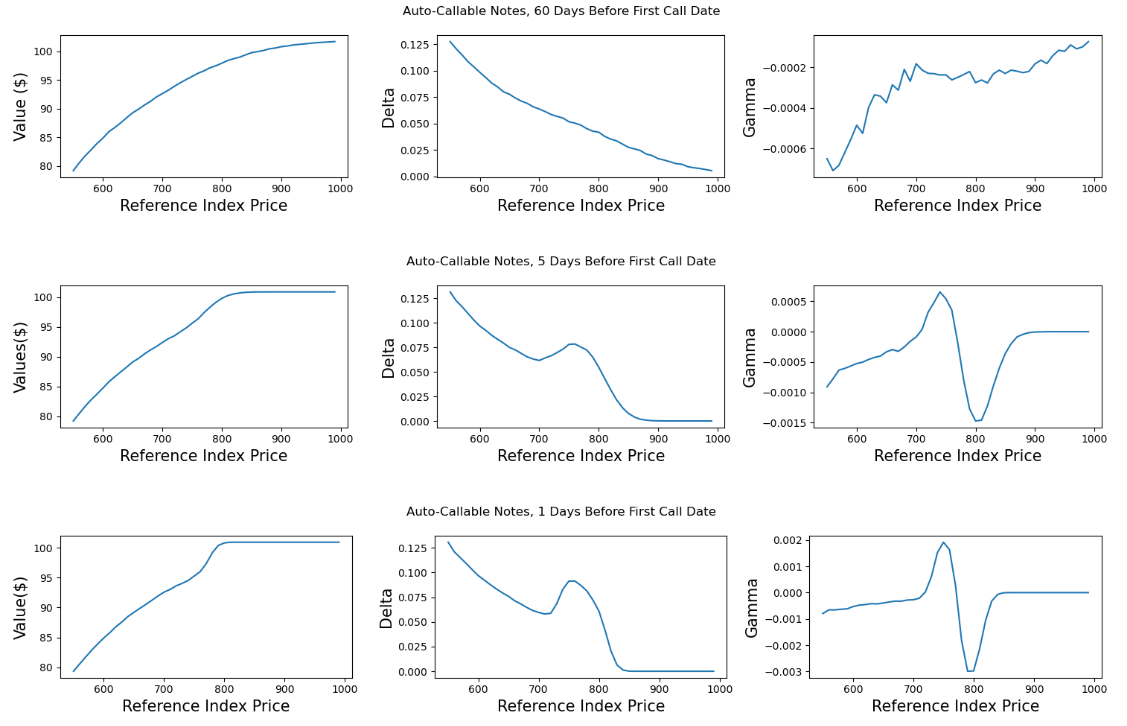}
\caption{Figures shows the value, $Delta$ and $Gamma$ of Autocallable notes at different underlying index price 60 days, 5 days and 1 day before the first call date.}
\label{fig:autocall_sensitivity}
\end{center}
\end{figure*}

% \textbf{PLEASE CHECK, terms are not in the table}
Autocallable notes are structured products that provide the investors with an opportunity to earn extra interest in terms of coupon payments if the underlying asset price closes above a specific threshold on periodic observation dates (barriers). In addition, the note will be autocalled or redeemed on an observation date if asset price return is above or equal to autocall barrier. Otherwise, it may offer contingent downside protection when the notes are held to maturity. 
%they provide autocall feature at each monitoring date prior to maturity. The Autocallable note is redeemed if the underlying asset price is above a predefined level (autocall barrier) and the investor receives both the principal amount and the coupon. 
Figure~\ref{fig:early_call_autocallable} shows this example when the note is auto called by the bank before the maturity of $7$ years. The note also provides principal protection at maturity if the Reference Index Return is greater than or equal to -35\% on the final valuation date. Table~\ref{tab:autocall_str} shows the note structure that we used in our experiments. We perform Monte Carlo simulation to simulate the note price. The note greeks ($Delta$ and $Gamma$) are generated using the finite difference method. Figure~\ref{fig:autocall_sensitivity} shows the variation of the option value, $Delta$, and the $Gamma$ value $1$ day, $5$ days and $60$ days before the call date. %The greeks value is very high near the call date and hence we choose a high rebalancing frequency. 
The note greeks change significantly near the observation date, making it difficult to maintain a Delta/Gamma neutral portfolio (as also highlighted in~\cite{autocall_hedge_mc}). 

% Figure~\ref{fig:autocall_sensitivity} shows the value, Delta and Gamma of Autocallable notes at different underlying index price 60 days, 5 days and 1 day before the first call date. The call price of the underlying index is 775. The automatic call feature introduces a heightened sensitivity in the Greeks as the call dates draw near, with particular emphasis on the call price. Close to the call dates, Delta exhibits a pronounced spike near the call price, and Gamma undergoes more drastic changes, assuming a sinusoidal shape. The intricate fluctuations in the Greeks make the hedging of Autocallable notes more complex compared to conventional derivatives. 

\subsection{Problem Formulation as an MDP}
The RL Pipeline uses a MDP (Markov Decision Process) to frame the problem of our RL agent interacting with the environment to maximize our potential reward. An MDP is defined as a tuple of elements ($S,\mathcal{A},f,R,\gamma$), where $S$ is the state space, $\mathcal{A}$ is the action space, $f(s_t,s_{t+1})$ is the state transition function, $R(s,a)$ is the reward function and $\gamma$ is the discount factor. We formulate a finite horizon discounted sum reward problem where the horizon length is the maturity of the option. 

\noindent The individual elements of the MDP are described below: 

\noindent\textbf{State:} States are the observations that the agent receives from the environment at each time step. In the current hedging pipeline, the state at time t is defined as $s_t = (x_t, \gamma_p, \tau)$, where $x_t$ is the stock price, $\gamma_p$ is the portfolio gamma, and $\tau$ is the time to next call date. %We use a different state vector when we use American/European options in trader/hedging portfolio.

\noindent\textbf{Action:} RL hedging agent’s action at any instant is the proportion of maximum hedging that can be done. For example, if the selected action is $0.2$, then the RL agent takes a position equal to the $20\%$ of the maximum hedge allowed. This proportion is then translated into actual number of units in the hedging instrument and those units are used to simulate the portfolio for the next time step. 

\noindent\textbf{Reward:} Reward is defined as the following:

\begin{equation}
\label{eq:reward}
R_i = -\kappa |V_i H_i| + (P_i^- - P_{i-1}^+)
\end{equation}

where $V_i$ is the value of the digital option at time $i\Delta t$, $H_i$ is the position taken in digital option, $\kappa$ is the transaction cost, $P_i^-$ is portfolio’s market value before time $i\Delta t$, $P_i^+$ is portfolio’s market value after time $i\Delta t$, $-\kappa |V_i H_i|$ is the transaction cost paid at time $i$, and $(P_i^- - P_{i-1}^+)$ is the change in portfolio value from time $(i-1)\Delta t$ to $i\Delta t$.

\noindent\textbf{State transition function:} With $s_t$ as the state at time $t$, the policy selects an action $a_t\in [0,1]$. The next state is updated based on the amount of hedging done at the current instant. The updated gamma of the portfolio and the new underlying price are then used in the state variable at the next instant. 

Using the above MDP, an RL environment was created which gives the next state of the environment based on the actions selected by the RL agent. At any given time, the RL agent interacts with the environment simulator by providing an action and the environment returns the next state and the reward.

\section{Proposed Method}
\label{sec:method}

We will now describe the RL method, it's architecture and the training procedure which learn the optimal policy.

% \begin{figure}[h]
% \includegraphics[width=8cm]{images/comb_pmf.JPG}
% \caption{The figure shows a PMF when the values of the distribution are equidistant.}
% \label{fig:comb_pmf}
% \end{figure}

\begin{figure}[h]
\centering
\includegraphics[width=8.3cm]{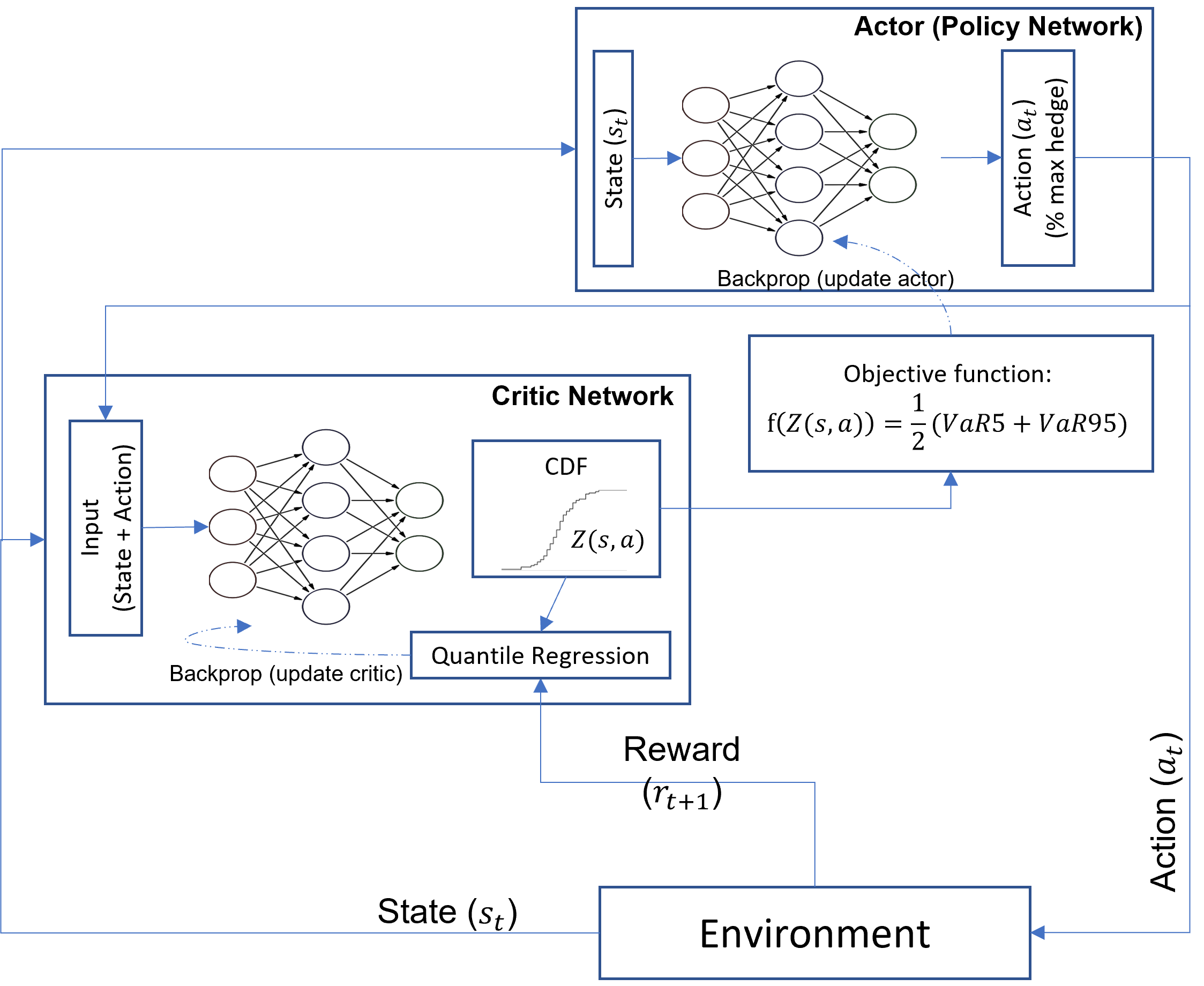}
\caption{Model architecture of Distributed Distributional DDPG (D4PG) with Quantile Regression (QR).}
\label{fig:d4pg}
\end{figure}

\begin{figure}[h]
\includegraphics[width=7cm]{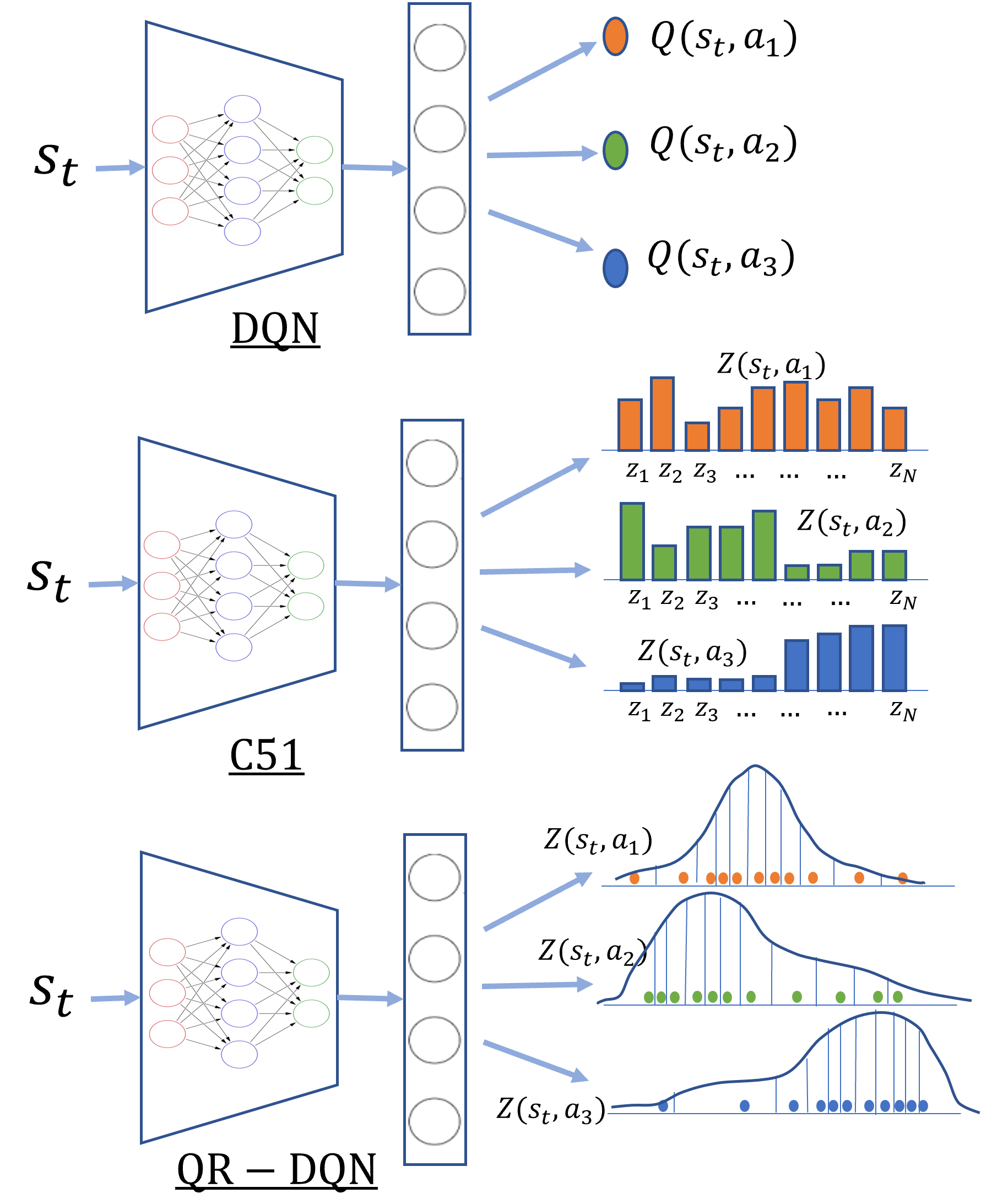}
\caption{Approaches to estimate distributions with DQN.}
\label{fig:c51_qr}
\end{figure}

~\subsection{Classical Reinforcement Learning}
In a sequential decision making process, reinforcement learning is found to make optimal decisions in the stochastically changing environment. At any time instant $t$, an RL agent takes an action $a_t$, and the environment, in response, returns the next state $s_{t+1}$ and the reward $r_{t+1}$ as a feedback of that action. The process continues until the maturity or maximum time reached~\cite{Book_Sutton}. The goal of the RL agent is to maximize the expected future reward at any time instant. The expected future reward $G$ at time $t$ (starting from state $s$ and taking action $a$) is: 
\begin{equation}
    G_t(s,a) = R_{t+1} + \gamma R_{t+2} + \gamma^2 R_{t+3} + \dots + \gamma^{T-1} R_{T}
\end{equation}

where $T$ is a horizon date and $\gamma\in (0,1]$ is a discount factor. $R_t$ is the reward received from the environment at time $t$.

Classical RL estimates the expected value of the future reward by learning a policy $\pi(s,a)$. $\pi(s,a): s\rightarrow a$ maps state $s$ to action $a$ for any time instant. If the current state is $s_t$ then the RL agent's action is determined as $a_t = \pi(s_t)$. The policy is iteratively learned using  reinforcement learning algorithms~\cite{Book_Sutton}, such as deep Q-learning (DQN), policy gradient, PPO, among others. 

Classical RL has shown tremendous improvement in several domains but in the realm of risk management, distributional RL is found to be more effective~\cite{cao2023_gamma_vega} to reduce the value-at-risk ($VaR$). We  utilize distributional RL~\cite{dabney2017distributional_qrdqn, d4pg_orig} to hedge the skewed distribution of the Autocallable structured note. 

~\subsection{Distributional Reinforcement Learning}
 We use Distributional Reinforcement Learning (DRL) in our proposal. DRL is an advanced paradigm which focuses on modeling the entire distribution of returns, rather than solely estimating the expected value. In DRL, the return $G_t$ is modelled as a distribution $Z_t^\pi$ for a fixed policy $\pi$. The return distribution provides more information and is more robust as compared to only the expectation. Classical RL tries to minimize the error between two expectations, expressed as  %$\mathop{\mathbb{E}}_{s,a,s^{'}}[(r(s,a) + \gamma max_{a^{'}} Q(s^{'},a^{'}) - Q(s,a))^2]$ 
$E_{s,a,s^{'}} [\{r(s,a) + \gamma max_{a^{'}} Q(s^{'},a^{'}) - Q(s,a)\}^2]$, 

where $Q(s,a)$ is the output of the policy and $r(s,a) + \gamma max_{a^{'}} Q(s^{'},a^{'})$ is the target function. In contrast, in DRL, the objective is to minimize a distributional error, which is a distance between full distributions~\cite{bellemare2017distributional_c51,dabney2017distributional_qrdqn}.
% \begin{equation}
%     sup_{s,a}  dist(R(s,a) + \gamma Z(s^{'},a^{*}),Z(s,a)), s^{'}\sim p(⋅|s,a)
% \end{equation}

The optimal action is selected from the $Q$ value function:
\begin{equation}
    % a^{*} = argmax_{a^{'}}{Q(s^{'},a^{'})} = argmax_{a^{'}} \E[Z(s^{'},a^{'})] 
    % a^* = \mathop{\mathbb{E}}[Z(s^',a^')] 
    a^{*} = \argmax_{a^{'}} Q(s^{'},a^{'}) = \argmax_{a^{'}}  E[Z(s^{'}, a^{'})]  
\end{equation}

In DRL, the distribution of returns is represented as a PMF (Probability Mass Function) and generally, the probabilities are assigned to discrete values that denote the possible outcome of the RL agent. Let’s say we have a neural network that predicts this PMF by taking a state $s$ and returning a distribution $Z(s,a)$ for each action. Categorical distributions are commonly employed to model these distributions in some DRL algorithms like C-51~\cite{bellemare2017distributional_c51} where the action distribution is modelled using a finite number of possible outcomes. In C-51, probabilities are estimated to these fixed locations. We employ Quantile Regression (QR) to learn the distribution of returns and unlike C-51, QR estimates the quantile locations where each quantile corresponds to a fixed uniform probability. That means, QR provides the flexibility to stochastically adjust the quantile locations in place of fixed locations in C-51. QR is a popular approach in DRL which is combined with several distributional RL algorithms such as in QR-DQN~\cite{dabney2017distributional_qrdqn}. QR-DQN uses quantile regression with traditional DQN to learn a distribution of outcomes. The goal of distributional DQN is that we want the distribution $Z(s,a)$ and target distribution $R_{t+1} + \gamma max_a Z(s_{t+1},a)$  as similar as possible, which is learned by minimizing the Wasserstein distance~\cite{dabney2017distributional_qrdqn} in the neural network model. Figure \ref{fig:c51_qr} illustrates the different variants to incorporate a distributional architecture with classical DQN. We have used QR with D4PG.

\subsubsection{Distributed Distributional DDPG (D4PG)}
We use the Deep Deterministic Policy Gradients (DDPG) algorithm to learn the underlying distribution of returns. DDPG is an Actor-Critic based method which helps to learn a policy in continuous action space. Specifically, we use D4PG (Distributed Distributional DDPG)~\cite{d4pg_orig} to learn the optimal policy. D4PG is a distributional RL algorithm which estimates the distribution of the return (unlike the mean in classical RL). Figure~\ref{fig:d4pg} shows the model architecture that we utilized in our experiments to train the D4PG algorithm. % \subsubsection{Network Architecture and Training}
There are three components:
% \subsubsection{Network Architecture}
\begin{enumerate}
    \item \textbf{The trading environment}: One essential component of the architecture is the trading environment, which simulates the stock and option prices. It tracks how the portfolio evolves over time based on the agent's hedging position, market dynamics (e.g., SABR volatility), and the arrival of the client options. At any particular time, the environment receives the hedging action and returns the next state and the reward.
    % it involves a simulator of how the portfolio composition/value evolves over time based on the hedging positions that the agent takes, the market dynamics for the underlying (e.g., SABR), and the client option arrival process. At any given time, the simulator computes the next state and the reward
    \item \textbf{The actor neural network} (also known as policy network) implements the hedging strategy. It is a neural network of size $(256, 256, 256, 1)$. %with \textit{Adam} optimizer
    At any instant $t$, it takes as input a state $s_t$ and outputs the amount of hedging ($a_t$) that the agent should perform. The objective function for training the agent's neural network is ½(5\% $VaR$+95\% $VaR$). % as the final objective function to train our agent’s policy network.  %We use a $4$ layer neural network with size $(256, 256, 256, 1)$ with \textit{Adam} optimizer. We use ½(5\%$VaR$+95\%$VaR$) as the final objective function to train our agent’s policy network. 
    \item \textbf{The critic neural network} takes as inputs a state, $s_t$, and the action from the actor’s output, $a_t$. Its role is to (a) estimate the distribution of the trading loss at the end of the hedging period, $Z(s_t, a_t)$, when taking action $a_t$ in state $s_t$, and (b) compute gradients that minimize the objective function $f(Z(s_t, a_t))$. We use a neural network of size $(512, 512, 256, 1)$ as the critic network. We use the reward from the environment to train the agent’s critic network. 

\end{enumerate}

We utilize quantile regression (QR)~\cite{dabney2017distributional_qrdqn} in combination with D4PG to approximate the distribution $Z(s, a)$ with the help of quantiles at the output of critic neural network. We use $100$ quantiles in our experiments for D4PG policy learning. Each quantile has a fixed probability but the location is stochastically adjusted during training. During agent-environment interaction in the training phase, an experience $(s,a,r,s^{'})$ is saved in a replay buffer which is repeated at every step. To update the policy, a batch of experiences are extracted from the replay buffer and %the policy output (i.e., the selected action) is moved towards the target distribution. 
the target distribution $r + \gamma Z(s^{'},a^{*})$ is used to compute the error from the policy output. Wasserstein distance~\cite{dabney2017distributional_qrdqn} is used as the loss function to estimate the quantiles. 

% \subsection{Network Architecture and Training}
% Explain loss function for both actor and critic, explain in brief about backpropagation. Figure~\ref{fig:d4pg}.

\section{Experiments and Results}
\label{sec:results}

In this section, we will describe the different empirical analysis that we performed on vanilla options and Autocallable notes, along with the experimental setup and the performance metrics. %First, we detail the experimental setup and the performance metrics. Secondly, we will show the performance of our reinforcement learning algorithm for hedging on both vanilla and Autocallable options. %Finally, we will show performance comparison with baselines strategies, Delta Neutral and Delta-Gamma Neutral. 

% \begin{figure*}[h]
%     % \includegraphics[width=8cm]{images/pnl_all_methods.jpg}
%     \includegraphics[width=18cm]{images/results_table_img.png}
% \caption{Performance comparison of baseline hedging strategies with RL hedging using American call/put and Digital options as hedging instruments.}
% \label{fig:perf_comparison_table}
% \end{figure*}

\begin{table*}[t]
\caption{The table shows performance comparison of the different hedging strategies with $2\%$ fee on a portfolio of American options. The (no Ex.) means that no early exercise was performed and (Early Ex.) signifies that the option was early exercised with Longstaff's method.}
\label{tab:perf_comp_american}
\centering
 \begin{tabular}{|c c c c c c c c c|}
\hline
 Strategy & Mean & Std & Mean-Std & 5\%$VaR$ & 5\%$CVaR$ & 95\%$VaR$ & 95\%$CVaR$ & Gamma Ratio\\ \hline
 Delta Neutral (no Ex.)  & 0.1 & 12.58 & -20.59 & 19.85 & -1.43 & -19.93 & -30.45 & 0.0 \\ 
 Del-Gamma N. (no Ex.) & -11.4 & 2.61 & -15.68 & -7.63 & -11.64 & -15.92 & -17.80 & 1.0 \\ 
 RL (no Ex.)  & -5.27 & 3.17 & -10.49 & -0.37 & -5.60 & -10.65 & -12.50 & 0.31 \\

 Delta Neutral (Early Ex.)  & 0.37 & 11.09 & -17.87 & 17.76 & -0.98 & -16.72 & -26.16 & 0.0 \\ 
 Del-Gamma N. (Early Ex.) & -11.76 & 2.71 & -16.22 & -7.84 & -12.01 & -16.47 & -18.40 & 1.0 \\ 
 RL (Early Ex.)  & -4.72 & 3.92 & -11.18 & 0.94 & -5.08 & -11.76 & -14.40 & 0.30 \\
 
 % RL [American Put \& Call] (1\% fee)  & -3.05 & 13.84 & -25.82 & -21.33 & -30.35 & 2.58 \\ 
 \hline
     
 \end{tabular}
\end{table*}

\subsection{Experimental Setup}
We are hedging a trader's portfolio which contains risky options - the trader hedges the portfolio by adding other instruments to a hedging portfolio at every hedging instant. In other words, the trader rebalances the portfolio at every $\Delta t$ time interval for hedging. % (eg., we use $\Delta t = 1$ month in our experiments for hedging Autocallable note). %At each rebalancing instant, an at-the-money American option or out-of-money Digital option is added to the hedging portfolio for hedging. 
The hedging action involves selecting the percentage of $Gamma$ to hedge at any rebalancing instant.
% The hedging action is to select the percentage of Gamma to hedge at any rebalancing instant. 
For example, in the experiments for Autocallable note hedging, the trader's portfolio contains one short Autocallable note with $7$ years of maturity. The trader also keeps a hedging portfolio and %The note structure and underlying are explained in section~\ref{sec:problem}. 
at every hedging instant ($\Delta t = 1$ month), the trader adds an at-the-money American call/put option or out-of-money Digital option to the hedging portfolio for hedging. A hedging strategy decides how much hedging needs to be performed based on the different portfolio Greeks such as $Delta$, $Gamma$, etc. $Delta$ tells how much the option's price will change for a one-point change in the underlying asset's price, while $Gamma$ tells how much the $Delta$ will change for a one-point change in the underlying asset's price. Trader's use these Greeks to manage risk and make informed decisions. For example, a Delta neutral strategies hedges the entire $Delta$ of the portfolio by creating a position with a $Delta$ value of zero, or very close to zero. %By constructing a delta-neutral portfolio, traders aim to minimize the impact of the underlying asset's price movements on the overall position. 
% In our experiments, the hedging action is to decide how much $Gamma$ needs to be hedged. %We use two baselines hedging strategies, Delta neutral and Delta-Gamma neutral. %For Delta neutral, the action is always $0$ (hedge only the Delta position) and for Delta-Gamma neutral, the action is always $1$ (minimize the whole Gamma). 
We propose a RL based hedging strategy which learns to decide the percentage of the $Gamma$ that needs to be hedged. We compare the hedging performance of our RL agent with traditional hedging strategies, Delta neutral and Delta-Gamma neutral. % and an existing RL algorithm which uses D4PG for vanilla European options~\cite{cao2023_gamma_vega}. 

%We will also compare the performance with an existing RL strategy as seen in~\cite{kolm_ritter,cao2023_gamma_vega}

\begin{figure}[h]
% \centering
\includegraphics[width=8cm]{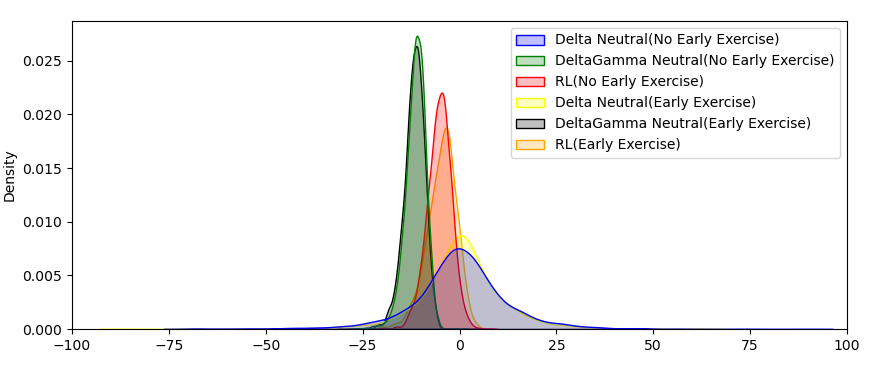}
\caption{The $PnL$ distribution of different hedging strategies on a portfolio containing American options.}
\label{fig:perf_comparison_american}
\end{figure}

% We show hedging of traditional strategies (Delta neutral and Delta-Gamma neutral), existing RL methods~\cite{kolm_ritter,cao2023_gamma_vega} and our proposed method using distributional RL. For Delta neutral, the action is always $0$ (hedge only the Delta position) and for Delta-Gamma neutral, the action is always $1$ (minimize the whole Gamma). 
% The action selection for existing RL algorithms is taken as they have proposed in their approach. 

Our RL agent uses the D4PG (Distributed Distributional DDPG) algorithm with quantile regression to learn an optimal policy. The RL agent selects an optimal action using the learned policy from interval $[0,1]$. The RL agent is trained for $40,000$ episodes (where one episode is one stock path till note maturity or auto-call). To compare the performance for all methods, we generate an additional $5000$ episodes on which KPI metrics are reported. The prices for the underlying asset are generated using Geometric Brownian motion (GBM) with a SABR volatility model. We evaluate the hedging performance using $95\%$ percentile of the value-at-risk ($VaR$) and conditional value-at-risk ($CVaR$) distribution. %All results reported in subsequent text are on the $5000$ evaluation episodes. 

% \noindent\textbf{Performance metric}: We evaluate the hedging performance using $95\%$ percentile of the value-at-risk (VaR) and conditional value-at-risk (CVaR) distribution.\\ %The value of risk is computed using 

% \begin{equation}
%     Value at Risk = v_m (v_i / v(i - 1))
% \end{equation}

\begin{table*}[h]
\caption{The table compares the performance of the Delta neutral, Delta-Gamma neutral, and RL agent on a portfolio of Autocallable notes with Digital and American options as hedging instruments with $2\%$ transaction cost.}
\label{tab:perf_comp}
\centering
 \begin{tabular}{|c c c c c c c c c|}
\hline
 Strategy & Mean & Std & Mean-Std & 5\%$VaR$ & 5\%$CVaR$ & 95\%$VaR$ & 95\%$CVaR$ & Gamma Ratio\\ \hline
 Delta Neutral  & -12.1 & 10.41 & -29.23 & -2.61 & -12.66 & -34.06 & -42.93 & 0.0 \\ 
 Delta-Gamma Neutral  & -10.15 & 8.99 & -24.94 & -1.21 & -10.69 & -28.12 & -36.13 & 1.0 \\  
 % [Digital] (0\% fee)  & -3.94 & 13.02 & -25.36 & -26.67 & -39.09 & 3.02 & - & -\\ 
 % RL (Gamma-Vega)   & -10.71 & 9.19 & -25.83 & -1.52 & -11.26 & -29.46 & -37.26 & 0.9 \\
 RL [Digital]  & -3.99 & 13.08 & -25.51 & 12.6 & -5.33 & -27.21 & -39.18 & 3.46 \\
 RL [Am. Put \& Call]   & -3.84 & 13.46 & -25.97  & 20.82 & -5.92 & -22.61 & -32.00 & 2.58 \\ 
 % RL [American Put \& Call] (1\% fee)  & -3.05 & 13.84 & -25.82 & -21.33 & -30.35 & 2.58 \\ 
 \hline
     
 \end{tabular}
\end{table*}

\noindent\textbf{Implementation:} We implemented the D4PG algorithm using the ACME library from Deepmind~\cite{hoffman2020acme} with Tensorflow backend. We utilized a server with $64$-GBs of RAM with $16$-CPU cores for training. To train the model, we employ a deep neural network with three hidden layers with \emph{Adam} optimizer for both actor and critic networks. The subsequent sections present the performance on the optimal parameters.

\subsection{RL Agent for Hedging }
We now describe the performance of the RL agent trained with a trader's portfolio containing client options. We first describe the performance of our RL agent on vanilla options where we achieve higher $PnL$ than baseline hedging strategies with simple training coupled with early exercise. We then describe the challenges in hedging a portfolio containing a structured product (Autocallable note) which contains several barriers and coupon payments. At the end, we show the performance comparison of the RL agent with traditional hedging strategies. \\

% We now describe the performance of our RL agent on a portfolio containing one short Autocallable note. We first describe in brief the performance on vanilla option which deliver good PnL with simple training, however, training the autocallable note was not straightforward due to the complex structure of Autocallable with several barriers and coupon payments. 

\noindent \textbf{Hedging vanilla Options:} In this experiment, we hedge a trader's portfolio containing American options where the portfolio is evolving regularly with time. The options are assumed to arrive as a Poisson process with $\lambda=1$. Each arrival option could be a call/put with long/short position with equal probability with a maturity of $1$ month. For this experiment, we use the portfolio structure as used by~\cite{cao2023_gamma_vega} to hedge a portfolio containing European options. Similar to European option, the American option also has a maturity of $1$ month but offers an extra feature to allow early exercise of the option before the maturity date. The method proposed by~\cite{cao2023_gamma_vega} doesn't support early exercise and we employ the Longstaff Schwartz algorithm~\cite{schwartz2001valuing_lsmc} to decide when to early exercise any American option in the portfolio. For hedging, American call and put options are used as hedging instruments. %The state at any instant $t$ contains the underlying stock price, portfolio gamma value, and the option gamma value. 
The hedging strategy decides how much hedging needs to be performed in terms of $Gamma$ value of the portfolio. The delta-neutral strategy hedges $0\%$ gamma (hedges only the $Delta$ exposure) and delta-gamma neutral strategy hedges $100\%$ gamma. The RL agent selects an action from $[0,1]$. %The RL agent is trained for $40,000$ episodes.

To early exercise the American option, we train the Longstaff Schwartz based regression method to predict the continuation value of the option given the stock path till date. The continuation value is the value of the option if it is not exercised at the current instant, while the exercise value is the payoff received when the option is executed at the current instant. For example. for a put option, the payoff $=max(K-x_t,0)$, where $K$ is the strike price and $x_t$ is the stock price at time $t$. 
% we predict the continuation value of the underlying price using the Longstaff-Schwartz method which is a backward iteration algorithm. 
We train the regression based algorithm on $10,000$ stock paths with several strike prices. At each instant during evaluation, the algorithm steps backward and approximates the continuation value of the stock. %The continuation value is the optimal payoff from continuing the holding for the current time-step. 
At any instant, if the current exercise value %(that is, for put option, the payoff $=max(K-x_t,0)$, where $K$ is the strike price and $x_t$ is the stock price at time $t$) 
is greater than the continuation value, then the option is exercised, else continued. This early exercise method is integrated with the RL method which uses D4PG with QR. Table~\ref{tab:perf_comp_american} shows the performance of different hedging strategies on the portfolio containing American options (without and with early exercise). Figure~\ref{fig:perf_comparison_american} shows the $PnL$ distribution of several hedging strategies, and shows significant improvement using our RL approach with early exercise of the American options (please note $5\% VaR$ in Table~\ref{tab:perf_comp_american}). In the figure, the distribution plot for the Delta neutral agent is quite wide but narrower for the Delta-Gamma neutral agent. Whereas, the $PnL$ distribution of the RL agent is a little wider than that with Delta-Gamma, and it is more on the positive side of the graph. The trend of the distribution for the RL agent on American options is that it achieves better results. The values in the table and the figure are on the $5000$ evaluation episodes.

% We assume that we have a trader's portfolio containing American options. The portfolio is evolving and the options arrive as a Poisson distribution and each arrival option could be a call/put with long/short position with equal probability. We follow a similar experiment setup for the vanilla American option as used by \cite{cao2023_gamma_vega} for European options. We used Longstaff Schwartz algorithm~\cite{schwartz2001valuing_lsmc} to decide early exercise of American options. We short American call/put options daily as hedging instrument to hedge the portfolio. Figure~\ref{fig:perf_comparison_american} shows the performance comparison of several hedging strategies which shows significant improvement using our RL approach with early exercise the American options. In the figure, the distribution plot for Delta agent is highly wide and narrower for the Delta-Gamma agent. Whereas, the RL agent is little wider than Delta-Gamma but is more on the positive side of the graph. The trend of the distribution for RL agent on American options achieve better results. 

\begin{figure}[h]
\centering
    \includegraphics[width=8cm]{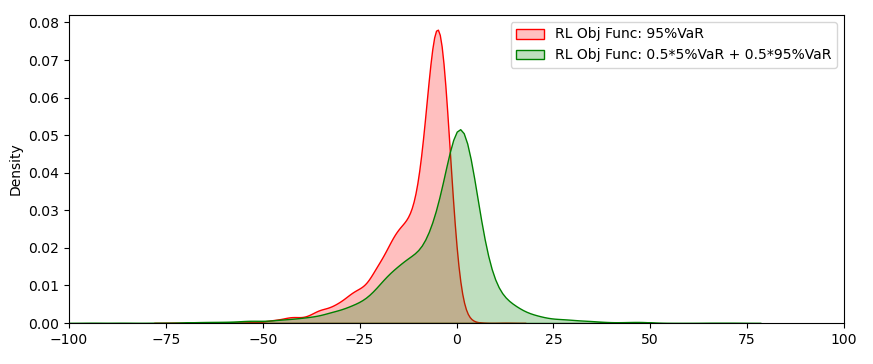}
\caption{The $PnL$ distribution on testing episodes when the RL agent when trained with the obj. function $95\% VaR$ and ($0.5*5\% VaR + 0.5*95\% VaR$).}
\label{fig:compare_obj_func}
\end{figure}

\begin{figure}[h]
\centering
    \includegraphics[width=8cm]{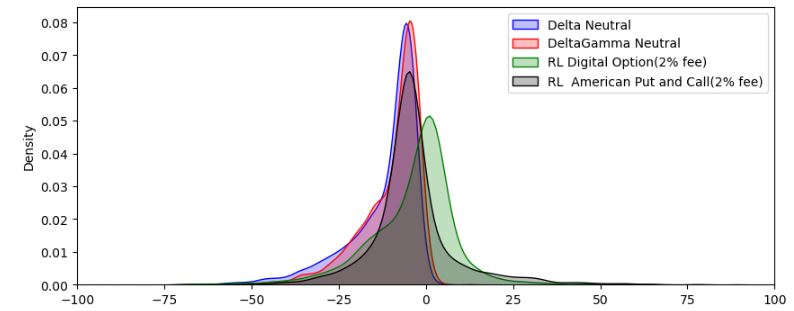}
\caption{The $PnL$ distribution of different hedging strategies.}
\label{fig:perf_comparison_autocall}
\end{figure}

\noindent\textbf{Hedging Autocallable structured note:}
In this experiment, we hedge a trader's portfolio containing one short Autocallable note with maturity of $7$ years. We run two experiments, one with a digital option as the hedging instrument and another using American call/put options. We use digital option as the hedging instrument because a digital option also contains a barrier similar to Autocallable notes but for a shorter maturity. For this experiment, the state ($s_t$) at any time instant $t$ contains the stock price ($x_t$), portfolio gamma ($\gamma_t$) and the days to the next call date ($\tau$). The trader rebalances the portfolio every month by adding one digital/American option in the hedging portfolio. We choose $\Delta t=1$ month as the rebalancing interval because of the longer maturity time of $7$ years for the Autocallable note. The hedging agent strategy decides how much hedging needs to be performed in terms of $Gamma$ value of the portfolio. The unhedged distribution for the Autocallable note is skewed as shown in figure~\ref{fig:pnl_no_hedge}. Unlike the vanilla option, the RL method that we applied for the vanilla options doesn't scale to this portfolio. We observed that the RL agent was not able to hedge the positive tail of the unhedged distribution. As a result, the RL agent selects only one action ($a=1$) irrespective of the state- i.e., it behaves like the Delta-Gamma neutral agent.  %that we have tried and the new objective function works better than the conventional objective function. 

Further investigation of this behavior during training, showed that the $95\% VaR $ as objective function makes the RL agent behaves similar to the delta-gamma agent. To add the positive $PnL$ in the objective function, we explored several variants and found experimentally that $0.5*5\%VaR + 0.5*95\%VaR$ is the objective function that assists to learn the optimal policy. Figure~\ref{fig:compare_obj_func} shows the comparison of the two objective functions $95\% VaR$ and ($0.5*5\% VaR + 0.5*95\% VaR$). We observed that the Autocallable barriers and regular coupon payments disturbed the reward function, leading to an inaccurate objective function at the critic. As a result, an optimal policy was not learned. The newly modified objective function generated the best payoff and the policy selected diverse actions. %We also hedged portfolio delta exposure using the underlying index. 
Figure~\ref{fig:perf_comparison_autocall} shows the $PnL$ distribution of our RL agent with the new objective function which learns the best distribution. The RL algorithm not only reduced $95\% VaR$ and increased $5\% VaR$ but also made the $PnL$ distribution more symmetric and shifted the distribution toward the more positive side. Table~\ref{tab:perf_comp} shows the $5\%VaR$, $95\% VaR$, mean and standard deviation of the $PnL$ over $5000$ evaluation episodes for the traditional strategies, namely delta-neutral, delta-gamma neutral, and our proposed RL method. %The results are reported for $5000$ evaluation episodes.

\section{Conclusion}

We proposed a reinforcement learning based method to hedge a portfolio of structured products such as Autocallable notes. The results showed that it is much more complex than the use of the traditional RL approach, which tends to fail. In this direction, we proposed a distributional RL based method to hedge a portfolio containing Autocallable structured notes.

% \appendix

% \section*{Ethical Statement}

% There are no ethical issues.

% \section*{Acknowledgments}

% The preparation of these instructions and the \LaTeX{} and Bib\TeX{}
% files that implement them was supported by Schlumberger Palo Alto
% Research, AT\&T Bell Laboratories, and Morgan Kaufmann Publishers.
% Preparation of the Microsoft Word file was supported by IJCAI.  An
% early version of this document was created by Shirley Jowell and Peter
% F. Patel-Schneider.  It was subsequently modified by Jennifer
% Ballentine, Thomas Dean, Bernhard Nebel, Daniel Pagenstecher,
% Kurt Steinkraus, Toby Walsh, Carles Sierra, Marc Pujol-Gonzalez,
% Francisco Cruz-Mencia and Edith Elkind.

\bibliographystyle{named}
\bibliography{ijcai24}

\begin{thebibliography}{}

\bibitem[\protect\citeauthoryear{Barth-Maron \bgroup \em et al.\egroup }{2018}]{d4pg_orig}
Gabriel Barth-Maron, Matthew~W. Hoffman, David Budden, Will Dabney, Dan Horgan, Dhruva TB, Alistair Muldal, Nicolas Heess, and Timothy Lillicrap.
\newblock Distributed distributional deterministic policy gradients, 2018.

\bibitem[\protect\citeauthoryear{Bellemare \bgroup \em et al.\egroup }{2017}]{bellemare2017distributional_c51}
Marc~G. Bellemare, Will Dabney, and Rémi Munos.
\newblock A distributional perspective on reinforcement learning, 2017.

\bibitem[\protect\citeauthoryear{Buehler \bgroup \em et al.\egroup }{2020}]{jpmorgan_market_friction}
Hans Buehler, Lukas Gonon, Josef Teichmann, Ben Wood, Baranidharan Mohan, and Jonathan Kochems.
\newblock Deep hedging: Hedging derivatives under generic market frictions using reinforcement learning.
\newblock {\em Available at SSRN}, 2020.

\bibitem[\protect\citeauthoryear{Cao \bgroup \em et al.\egroup }{2020}]{2q_Cao_2020}
Jay Cao, Jacky Chen, John Hull, and Zissis Poulos.
\newblock Deep hedging of derivatives using reinforcement learning.
\newblock {\em The Journal of Financial Data Science}, 3(1):10–27, December 2020.

\bibitem[\protect\citeauthoryear{Cao \bgroup \em et al.\egroup }{2023}]{cao2023_gamma_vega}
Jay Cao, Jacky Chen, Soroush Farghadani, John Hull, Zissis Poulos, Zeyu Wang, and Jun Yuan.
\newblock Gamma and vega hedging using deep distributional reinforcement learning, 2023.

\bibitem[\protect\citeauthoryear{Chen \bgroup \em et al.\egroup }{2023}]{chen2023hedging_barrier}
Jacky Chen, Yu~Fu, John~C Hull, Zissis Poulos, Zeyu Wang, and Jun Yuan.
\newblock Hedging barrier options using reinforcement learning.
\newblock 2023.

\bibitem[\protect\citeauthoryear{Cui \bgroup \em et al.\egroup }{2023}]{autocall_hedge_mc}
Yeda Cui, Lingfei Li, and Gongqiu Zhang.
\newblock Pricing and hedging autocallable products by markov chain approximation.
\newblock {\em Available at SSRN 4557397}, 2023.

\bibitem[\protect\citeauthoryear{Dabney \bgroup \em et al.\egroup }{2017}]{dabney2017distributional_qrdqn}
Will Dabney, Mark Rowland, Marc~G. Bellemare, and Rémi Munos.
\newblock Distributional reinforcement learning with quantile regression, 2017.

\bibitem[\protect\citeauthoryear{Daluiso \bgroup \em et al.\egroup }{2023}]{cva_hedge}
Roberto Daluiso, Marco Pinciroli, Michele Trapletti, and Edoardo Vittori.
\newblock Cva hedging with reinforcement learning.
\newblock In {\em Proceedings of the Fourth ACM International Conference on AI in Finance}, ICAIF '23, page 261–269, New York, NY, USA, 2023. Association for Computing Machinery.

\bibitem[\protect\citeauthoryear{Ganesh \bgroup \em et al.\egroup }{2019}]{ganesh2019reinforcement_marketmaking}
Sumitra Ganesh, Nelson Vadori, Mengda Xu, Hua Zheng, Prashant Reddy, and Manuela Veloso.
\newblock Reinforcement learning for market making in a multi-agent dealer market, 2019.

\bibitem[\protect\citeauthoryear{Gao \bgroup \em et al.\egroup }{2023}]{deeper_hedging_Gao_2023}
Kang Gao, Stephen Weston, Perukrishnen Vytelingum, Namid Stillman, Wayne Luk, and Ce~Guo.
\newblock Deeper hedging: A new agent-based model for effective deep hedging.
\newblock In {\em 4th ACM International Conference on AI in Finance}, ICAIF ’23. ACM, November 2023.

\bibitem[\protect\citeauthoryear{Giurca and Borovkova}{2021}]{2021_giurca_delta}
Alexandru Giurca and Svetlana Borovkova.
\newblock Delta hedging of derivatives using deep reinforcement learning.
\newblock {\em Available at SSRN 3847272}, 2021.

\bibitem[\protect\citeauthoryear{Halperin}{2019}]{halperin2019qlbs}
Igor Halperin.
\newblock Qlbs: Q-learner in the black-scholes(-merton) worlds, 2019.

\bibitem[\protect\citeauthoryear{Hoffman \bgroup \em et al.\egroup }{2020}]{hoffman2020acme}
Matthew~W. Hoffman, Bobak Shahriari, John Aslanides, Gabriel Barth-Maron, Nikola Momchev, Danila Sinopalnikov, Piotr Sta\'nczyk, Sabela Ramos, Anton Raichuk, Damien Vincent, L\'eonard Hussenot, Robert Dadashi, Gabriel Dulac-Arnold, Manu Orsini, Alexis Jacq, Johan Ferret, Nino Vieillard, Seyed Kamyar~Seyed Ghasemipour, Sertan Girgin, Olivier Pietquin, Feryal Behbahani, Tamara Norman, Abbas Abdolmaleki, Albin Cassirer, Fan Yang, Kate Baumli, Sarah Henderson, Abe Friesen, Ruba Haroun, Alex Novikov, Sergio~G\'omez Colmenarejo, Serkan Cabi, Caglar Gulcehre, Tom~Le Paine, Srivatsan Srinivasan, Andrew Cowie, Ziyu Wang, Bilal Piot, and Nando de~Freitas.
\newblock Acme: A research framework for distributed reinforcement learning.
\newblock {\em arXiv preprint arXiv:2006.00979}, 2020.

\bibitem[\protect\citeauthoryear{Khraishi and Okhrati}{2022}]{khraishi2022offline_creditpricing_icaif22}
Raad Khraishi and Ramin Okhrati.
\newblock Offline deep reinforcement learning for dynamic pricing of consumer credit, 2022.

\bibitem[\protect\citeauthoryear{Kolm and Ritter}{2019}]{kolm_ritter}
Petter~N Kolm and Gordon Ritter.
\newblock Dynamic replication and hedging: A reinforcement learning approach.
\newblock {\em The Journal of Financial Data Science}, 1(1):159--171, 2019.

\bibitem[\protect\citeauthoryear{Li \bgroup \em et al.\egroup }{2009}]{pmlr-v5-li09d_earlyexerciseRL}
Yuxi Li, Csaba Szepesvari, and Dale Schuurmans.
\newblock Learning exercise policies for american options.
\newblock In David van Dyk and Max Welling, editors, {\em Proceedings of the Twelth International Conference on Artificial Intelligence and Statistics}, volume~5 of {\em Proceedings of Machine Learning Research}, pages 352--359, Hilton Clearwater Beach Resort, Clearwater Beach, Florida USA, 16--18 Apr 2009. PMLR.

\bibitem[\protect\citeauthoryear{Mikkilä and Kanniainen}{2023}]{2022_empirical_deep_hedging}
Oskari Mikkilä and Juho Kanniainen.
\newblock Empirical deep hedging.
\newblock {\em Quantitative Finance}, 23(1):111--122, 2023.

\bibitem[\protect\citeauthoryear{Murray \bgroup \em et al.\egroup }{2022}]{murray2022deep_icaif22}
Phillip Murray, Ben Wood, Hans Buehler, Magnus Wiese, and Mikko~S. Pakkanen.
\newblock Deep hedging: Continuous reinforcement learning for hedging of general portfolios across multiple risk aversions, 2022.

\bibitem[\protect\citeauthoryear{Schwartz}{2001}]{schwartz2001valuing_lsmc}
Francis A Longstaff Eduardo~S Schwartz.
\newblock Valuing american options by simulation: A simple least—squares.
\newblock 2001.

\bibitem[\protect\citeauthoryear{Sutton and Barto}{1998}]{Book_Sutton}
Richard~S. Sutton and Andrew~G. Barto.
\newblock {\em Introduction to Reinforcement Learning}.
\newblock MIT Press, Cambridge, MA, USA, 1st edition, 1998.

\bibitem[\protect\citeauthoryear{Zhang \bgroup \em et al.\egroup }{2023}]{zhang2023generalizable_optimaltrade}
Chuheng Zhang, Yitong Duan, Xiaoyu Chen, Jianyu Chen, Jian Li, and Li~Zhao.
\newblock Towards generalizable reinforcement learning for trade execution, 2023.

\end{thebibliography}

\end{document}